\documentclass[12pt,english,aps,prl,twocolumn,showpacs,superscriptaddress,footinbib,reprint,noshowpacs]{revtex4-1}
\usepackage{color}
\usepackage{graphicx}
\usepackage{amsmath}
\usepackage{comment}
\usepackage{amssymb}

\usepackage{subfigure}

\begin{document}

\title{Geometry of energy landscapes and the optimizability of deep neural networks}

\author{Simon Becker}
\affiliation{Department of Applied Mathematics and Theoretical Physics, University of Cambridge, Wilberforce Road, Cambridge, CB3 0WA, United Kingdom}

\author{Yao Zhang}
\affiliation{Cavendish Laboratory, University of Cambridge, Cambridge CB3 0HE, United Kingdom}

\author{Alpha A. Lee}
\email{aal44@cam.ac.uk}
\affiliation{Cavendish Laboratory, University of Cambridge, Cambridge CB3 0HE, United Kingdom}

\begin{abstract}
Deep neural networks are workhorse models in machine learning with multiple layers of non-linear functions composed in series. Their loss function is highly non-convex, yet  empirically even gradient descent minimisation is sufficient to arrive at accurate and predictive models. It is hitherto unknown why are deep neural networks easily optimizable. We analyze the energy landscape of a spin glass model of deep neural networks using random matrix theory and algebraic geometry. We analytically show that the multilayered structure holds the key to optimizability: Fixing the number of parameters and increasing network depth, the number of stationary points in the loss function decreases, minima become more clustered in parameter space, and the tradeoff between the depth and width of minima becomes less severe. Our analytical results are numerically verified through comparison with neural networks trained on a set of classical benchmark datasets. Our model uncovers generic design principles of machine learning models.  
\end{abstract} 

\maketitle
\makeatother

Nonlinear multiparameter fitting is an ubiquitous in science, from cosmology \cite{leclercq2014cosmology} to biophysics \cite{gutenkunst2007universally}. The key challenge is non-convexity: Typically fitting is done by finding parameters that minimise the discrepancy between model prediction and data, known as the loss function. The loss function of non-linear models often have many minima and minimisation algorithms converge to local minima rather than the global minimum. 

Nonetheless, models often used in machine learning appear to circumvent this problem. The workhorse model, deep neural networks \cite{lecun2015deep}, comprises multiple layers of non-linear functions composed in series.  Deep neural networks achieved near-human accuracy in tasks such as image recognition \cite{krizhevsky2012imagenet} and translation \cite{wu2016google}. However, the success of deep neural network raises two fundamental unsolved puzzles: First, industrial models have millions of parameters \cite{russakovsky2015imagenet} and the loss function is highly non-convex, yet surprisingly even simple gradient descent algorithm is able to find accurate and predictive models. Second, it is long known that ``shallow'' neural networks -- models that comprise a sum, rather than composition, of non-linear functions -- can approximate any smooth function \cite{cybenko1989approximation}. However, deep neural networks empirically outperform shallower neural networks \cite{goodfellow2016deep}.   

The surprising effectiveness of deep neural networks is often explained in terms of the classes of expressible functions. Seminal works show that the multilayered structure allows deep neural networks to disentangle highly curved manifolds in input space into flat manifolds \cite{montufar2014number,poole2016exponential,poggio2017and}. Some argue that deep neural networks expresses ``physical'' functions: they can be mapped to the renormalisation group \cite{mehta2014exact} and implicitly imposes the physics of symmetry, locality and compositionality \cite{lin2017does}. However, recent numerical experiments problematize explanations based expressivity: shallower neural networks can match the accuracy of deep neural networks as long as the trained deep neural network is used augment the dataset by predicting labels of unlabelled data \cite{ba2014deep}. This observation suggests that deep and shallow networks are comparable in expressivity.  Explanation of why deep neural networks are effective must therefore turn to whether one can actually find optimal parameters given data, i.e. optimisability. 

Pioneering works show that for Gaussian random functions, critical points that take a value much larger than the global minimum are exponentially likely to be saddle points in the high dimensional limit  \cite{fyodorov2004complexity,bray2007statistics,fyodorov2007replica,auffinger2013complexity,auffinger2013random}. Modelling a neural network as a Gaussian random function, some argue that the value of the loss function at most local minima is similar to the global minimum and this is why local minima are ``good enough'' \cite{dauphin2014identifying,choromanska2015loss,choromanska2015open}. However, this does not directly explain why deep neural networks, in particular, outperform shallow neural network. Pioneering and seminal numerical studies of the energy landscape of loss functions using methods developed for molecular systems \cite{das2016energy,ballard2016energy,ballard2017energy,mehta2018loss} focused on shallow neural networks. 

In this Letter, we build on the spin glass model of deep neural networks introduced in \cite{choromanska2015loss} and derive novel analytical results describing the geometry of the loss function landscape as a function of network depth. We show that fixing the number of parameters and increasing network depth, the number of stationary points in the loss function decreases, minima become more clustered in parameter space, and the tradeoff between the depth and width of minima becomes less severe. We verify our results through comparison with neural networks trained on a set of classical benchmark datasets. 

We consider a fully connected feed-forward network with $H-1$ hidden layers where layer $k-1$ has $n_{k-1}$ nodes and each of them is connected to the $n_{k}$ nodes of layer $k$. The networks we consider take input vectors $\mathbf{X} \in \mathbb{R}^{n_0}$ entering the $0$-th layer and returns scalar outputs $Y$ from the $H$-th layer
\begin{equation}
Y(\mathbf{X},\mathbf{w})=q\theta(\mathbf{W}_H^T\theta(\mathbf{W}_{H-1}^T...\theta(\mathbf{W}_1^T \mathbf{X}))) 
\label{eq:neuralnet}
\end{equation} 
where the matrices $\mathbf{W}_k$ contain the weights $\mathbf{w}$ and the functions $\theta$ are the activation functions. We restrict the analysis to the commonly used rectified linear units ($\operatorname{ReLUs}$) $\theta(x)=\operatorname{max}(x,0)$. The normalising constant $q$ will be specified later to compare different architectures.
We label paths in the network as $(i,j)$ where $j$ labels any of the $P$ paths from a given component $X_i$ of the input vector. The quantity $w_{(i,j)}^{(k)}$ denotes the weight connecting layer $k-1$ with layer $k$ along path $(i,j)$. 

For simplicity, we consider a classification task: Let $\zeta=\operatorname{max}_{w}\left\lvert Y(\mathbf X,\mathbf{w}) \right\rvert$ be the maximum of the absolute value of the network output for admissible weight configurations. We consider a random labelling scenario where the ground truth $Y_{\text{true}}$ takes values $\pm \zeta$ independent of input $\mathbf{X}$. Our goal is to characterise the loss function $ \mathcal{L}(\mathbf w) = \mathbb E_A \left\lvert Y_{\text{true}}-Y(\mathbf X, \mathbf w) \right\rvert$ for this randomly labelled dataset. 

To make analytical progress, we map this neural network architecture onto a spin glass Hamiltonian via a series of elegant approximations introduced in \cite{choromanska2015loss}.  We rewrite \eqref{eq:neuralnet} by replacing the ReLUs by activation functions $A\in \{0,1\}$, 
\begin{equation}
Y(\mathbf{X},\mathbf{w})= q\sum_{i=1}^{n_0} \sum_{j=1}^{P} X_{i} A_{(i,j)} \prod_{k=1}^H w_{(i,j)}^{(k)}.
\label{eq:pathoutput}
\end{equation}

\begin{figure}
  \centering
   \includegraphics[width=8cm]{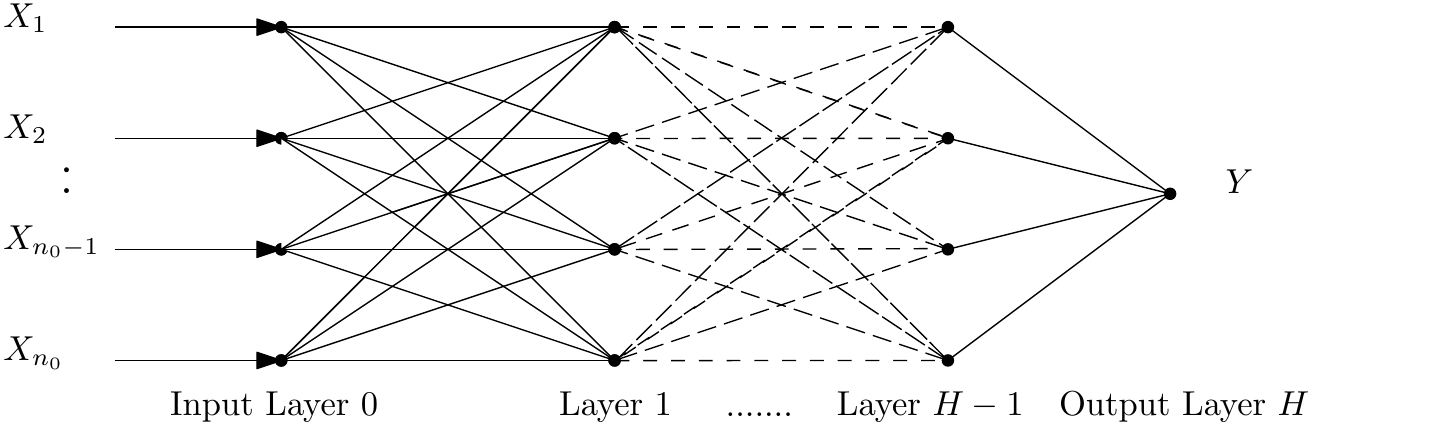} 
   \caption{\label{fig:Netarch0} A schematic of the the feedforward network architecture studied in this paper.}
\end{figure}

We next introduce the key approximations: First, the input of the network is assumed to consist of independent and standard normally distributed random variables. The activation functions $A$ are independent and Bernoulli distributed with probability $p$ of being 1. Second, the number of different weights $\Lambda$ is assumed to be the $H$-th root of the total number of paths in the network. Moreover, among all possible weight combinations of the $\Lambda$ number of weights, each configuration is assumed to appear almost equally often. Third, the weights $(w_n)$ are assumed to satisfy, after rescaling, a spherical constraint $\frac{1}{\Lambda} \sum_{n=1}^{\Lambda} w_n^2=1$. This spherical constraint models regularisation methods commonly used in the literature that penalises the magnitude of the weights. 

Under the three previously stated assumptions, and choosing $q=\Lambda^{-(H-1)/2}$, the loss function $\mathcal{L}(\mathbf w)$ has the same distribution as $p  \mathcal{H}_{\lambda}(\mathbf{w})$, where $\mathcal{H}_{\lambda}(\mathbf{w})$ is the $H$-spin spherical spin glass Hamiltonian 
\begin{equation}
 \mathcal{H}_{\Lambda}(\mathbf{w}) = \frac{1}{\Lambda^{(H-1)/2}} \sum_{i_1,..,i_H=1}^{\Lambda}Z_{i_1,...,i_H} \prod_{k=1}^H w_{i_k}  
 \label{eq:spinglass}
\end{equation}
and $Z_{i_1,...,i_H}$ are independent, identical, and standard normally distributed. 

We consider networks with different number of layers $H$ but with the same number of parameters $N_e$. All layers aside from the scalar output layer shall be assumed to be of equal size $n_0=...=n_{H-1}$ as in Fig. \ref{fig:Netarch0}. The number of network parameters $N_e = (H-1)n_0^2+n_0$ and 
\begin{equation}
\Lambda= \frac{\sqrt{4N_e(H-1)+1}+1}{2(H-1)}.
\label{eq:constraint}
\end{equation}

\emph{Number of critical points}: The spin glass Hamiltonian (\ref{eq:spinglass}) is evidently non-convex. Thus a natural question to ask is how does the number of critical point varies as the function of number of layers. The number of critical points $\mathcal{N}$ satisfies a remarkably simple theorem
\begin{equation}
\mathcal{N}=\frac{(H-1)^{\Lambda}-1}{H-2}. 
\label{eq:numcritweights}
\end{equation}
%

\emph{Proof}:  The loss function can be represented by a homogeneous symmetric random polynomial. To fix ideas we illustrate the link between the two for $H=2$ when the Hamiltonian is just $\mathcal{H}_{\Lambda}(\mathbf{w}) =  \sum_{i_1=1}^{\Lambda}\tfrac{X_{i_1,i_1}}{\sqrt{\Lambda}}w_{i_1}^2+ \sum_{i_1<i_2}^{\Lambda}\tfrac{(X_{i_1,i_2}+X_{i_2,i_1})}{\sqrt{\Lambda}}w_{i_1}w_{i_2}.$ In order to have a sum of random variables $Y_{i_1,i_2}+Y_{i_2,i_1}$ with the symmetry property $Y_{i_1,i_2}=Y_{i_2,i_1}$ to be distributed like $X_{i_1,i_2}+X_{i_2,i_1}$ one can choose $Y_{i_1,i_2} = \tfrac{X_{i_1,i_2}+X_{i_2,i_1}}{2} \sim \mathcal N(0,1/2).$ Critical weights $\mathbf{w}$ of $\mathcal{H}_{\Lambda}(\mathbf{w})$ are precisely the generalized eigenvectors satisfying for $j \in \left\{1,..,\Lambda\right\}$ the eigenvalue equation $\frac{1}{\Lambda^{(H-1)/2}} \sum_{i_2,..,i_H=1}^{\Lambda}Y_{\mathbf{j},...,i_H} \prod_{k=2}^H w_{i_k} = \lambda w_{\mathbf{j}}$ where two solutions $(\lambda,\mathbf{w}),(\lambda',\mathbf{w}')$ to the eigenproblem coincide if there is $t \neq 0$ such that $t\lambda^{H-2}=\lambda'$ and $t \mathbf{w}=\mathbf{w}'.$ Substituting $\lambda=\gamma^{H-2}$ in the eigenvalue equation yields $\Lambda$-many homogeneous equations of degree $H-1$ in $\Lambda+1$ many variables $\lambda, w_1,..,w_{\Lambda}$. The multi-homogeneous B\'ezout's theorem \cite[Ch. $4$, Sec. $2.2$]{Shafarevich1977} implies that such an equation has exactly $(H-1)^{\Lambda}$ solutions where we discard the equivalence class of the zero solution $\lambda=w_i=0$ to end up with  $(H-1)^{\Lambda}-1$ solutions. Removing the $H-2$ degeneracy, due to roots of unity $e^{2\pi i/(H-2)},$ coming from the $\lambda=\gamma^{H-2}$-substitution, shows that the number of critical weights satisfies Equation (\ref{eq:numcritweights}). This has been obtained using methods from toric geometry in \cite[Theorem $1.2$]{CS} (see Supplemental Materials (SM)).

\begin{figure}
  \centering
   \includegraphics[width=7cm]{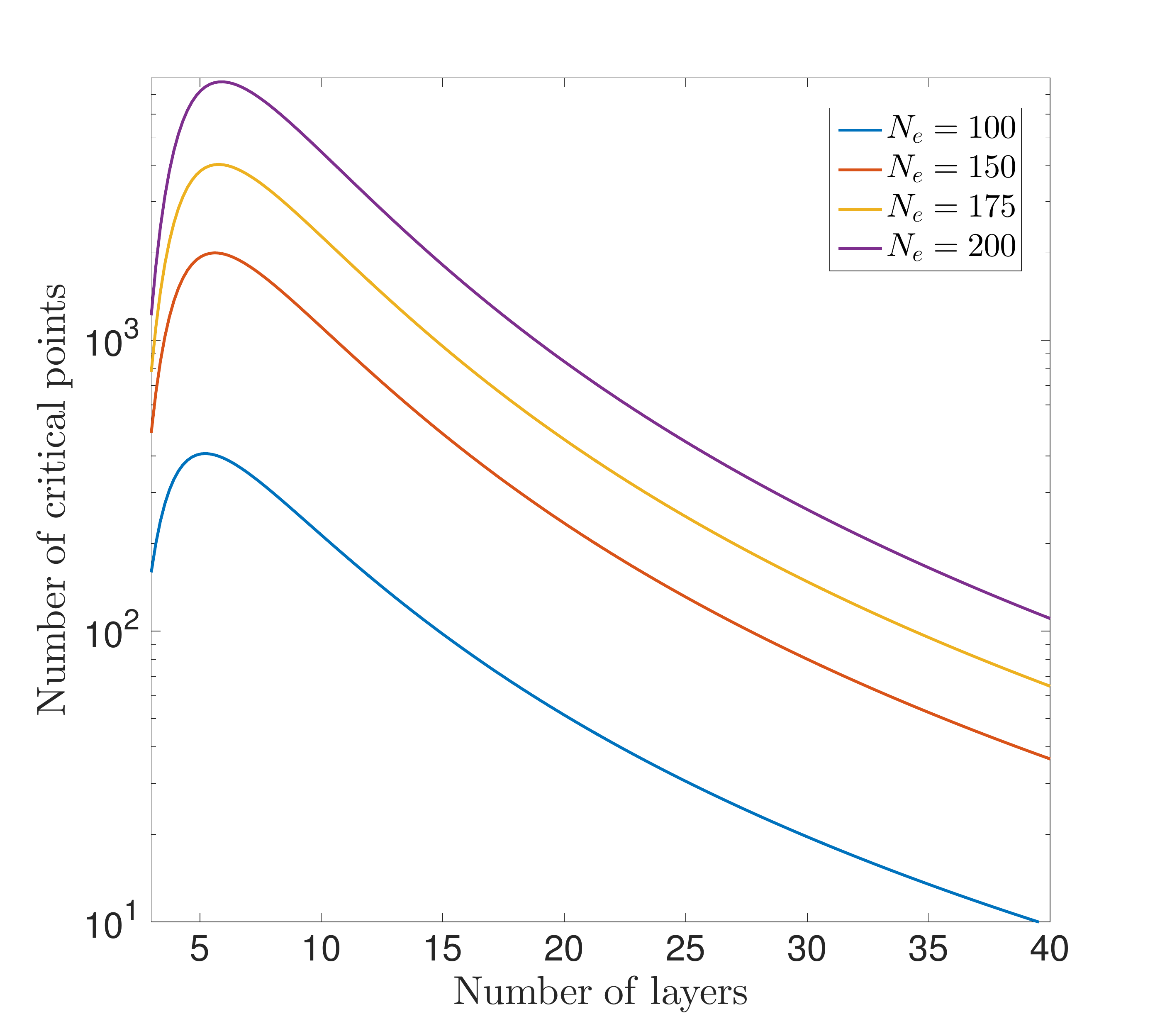} 
   \caption{The number of critical points $\mathcal{N}$ in a deep neural network decreases as a function of depth for fixed number of parameters $N_e$. }
   \label{Crit_points}
\end{figure}

Figure \ref{Crit_points} show that Equation (\ref{eq:numcritweights}) implies that the number of critical points is a non-monotonic function of the number of layers. Importantly, the number of critical points decreases as the number of layers increases for a deep network, thus deep networks are more optimisable because there are less critical points that traps the optimiser. Figure \ref{Crit_points} also shows that the number of critical points increases as a function of depth for shallow networks. This agrees with the early experience with deep learning in the 1980s and 1990s -- a one layer neural network is inefficient in learning compositional features, yet simply adding a few more layers to a one layer neural network causes performance to deteriorate because the number of critical points proliferates and the loss function becomes non-optimisable \cite{goodfellow2016deep}. The deep learning boon began when there were sufficient computational resources to train a very deep neural network. 

\emph{Location of minima}: Having considered how many critical points are there in a deep neural network, we next consider where are critical points located in weight space. Intuitively, the more clustered they are, the easier it is for an optimiser to search for minima. 
Let $\text{Crt}(-\infty, \mathcal E)$ denote the set of critical points for which the loss function takes values in $(-\infty,\Lambda \mathcal E).$ For an interval $I \subset[-1,1]$ we study the number of pairs $(\mathbf{w},\mathbf{w}')$ of critical weights in $\text{Crt}(-\infty, \mathcal E)$ with relative angle $\mathbf{w} \cdot \mathbf{w}'/\Lambda$ contained in $I.$ This set will be denoted by $\left[ \text{Crt}((-\infty, \mathcal E), I)\right]_2$. Note that the Euclidean distance $\left\lVert \mathbf{w} - \mathbf{w}' \right\rVert_2= \sqrt{2(\Lambda-\mathbf{w} \cdot \mathbf{w}')}.$ As we study large $\Lambda$-asymptotics, minima occur predominantly at low energies such that we may assume that all energies are sufficiently small, i.e. $\mathcal E/p \in (-\infty, -\sqrt{2}/\sigma]$ where $\sigma=\sqrt{H/(2(H-1))}$.

Our second theorem is that upper bound to distance between minima is 
\begin{equation}
\begin{split}
&\limsup_{\Lambda \rightarrow \infty} \frac{1}{\Lambda} \log\left(\frac{\mathbb E \left\lvert\left[ \operatorname{Crt}((-\infty, \mathcal E) , I)\right]_2\right\rvert}{\mathbb E \left\lvert \operatorname{Crt}(-\infty, \mathcal E) \right\rvert}\right) \\
&\le \sup_{r \in I} \sup_{v \in (-\infty, \mathcal E/p) } \Psi_H(r,v,\mathcal E)
\label{eq:relpos}
\end{split}
\end{equation}
where 
\begin{equation*}
\begin{split}
&\Psi_H(r,v,\mathcal E)=  \tfrac{1}{2} +\tfrac{\mathcal E^2}{2 p^2}+\tfrac{1}{2} \log\left( \tfrac{(H-1)(1-r^2)}{1-r^{2H-2}}\right)\\
&-\frac{1}{2}\left\langle \left(\begin{matrix} v \\ v \end{matrix}\right) , \Sigma_U(r)^{-1}\left(\begin{matrix} v \\ v \end{matrix} \right)\right\rangle + \int_{-2}^{2} \tfrac{ \log \left\lvert \sqrt{2} \sigma v-x \right\rvert\sqrt{4-x^2}}{2\pi} \ \mathrm{d} x.
\end{split}
\end{equation*} 
$\Sigma(r) = -\tfrac{1}{H} \left( \begin{matrix} b_1(r) & b_2(r) \\ b_2(r) & b_1(r) \end{matrix}\right)$ is a matrix defined by 
\begin{equation*}
\begin{split}
\alpha_H(r)&=(H-H(r^H-(H-1)(r^{H-2}-r^{H}))^2)^{-1},\\ b_1(r)&=-H+\alpha_H(r)H^3 (r^{2H-2}-r^{2H})\text{, and }\\
b_2(r)&= -Hr^H-\alpha_H(r)H^3r^{3H-4}(r^2+H(r^2-1)^2-1).
\end{split}
\end{equation*}

\emph{Proof}: The full proof is in the SM. Our proof strategy combines the asymptotics for the minima of the Hamiltonian \cite[Theorem $10$]{S17} with the upper bound on the angle between minima \cite[Theorem $5$ and Lemma $6$]{S17}. 

\begin{figure}
  \centering
   \includegraphics[width=7cm]{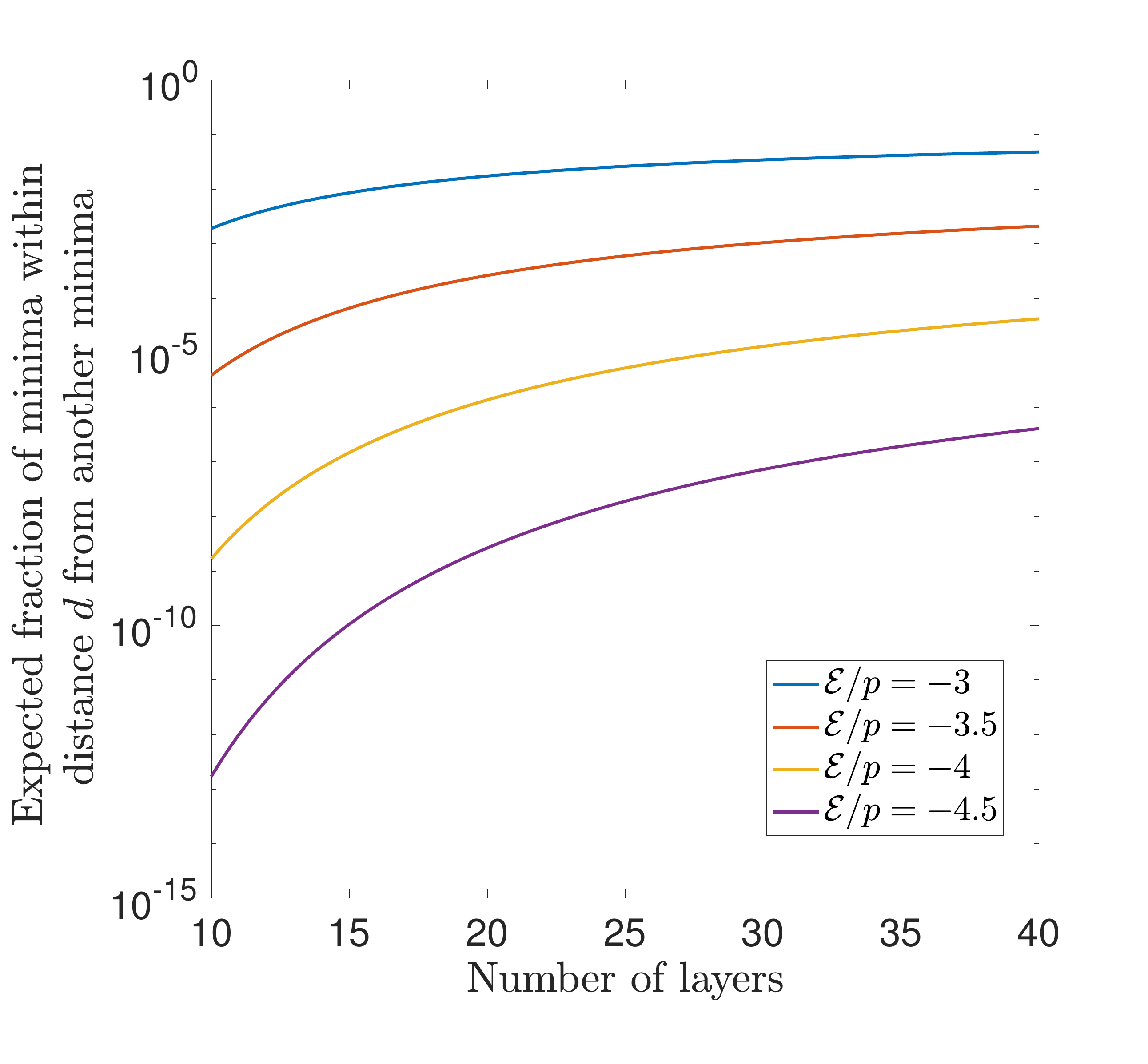} 
   \caption{Minima are more clustered for deeper networks. \label{fig:Netarch3} The figure shows the relative expected number of critical points \eqref{eq:relpos} that attains a loss function value in the interval $(-\infty, \Lambda \mathcal{E})$ with $||\mathbf{w}- \mathbf{w}'||_2^2  \le d \Lambda$ with $d=0.02$ for fixed number of network parameters $N_e=400$. }
\label{dist_crit_angle}
\end{figure}

Figure \ref{dist_crit_angle} shows that the number of minima, relative to the total number of minima, that are close to other minima (c.f. Equation (\ref{eq:relpos})) increases as the number of layers increases. In other words, minima are more clustered for deeper networks, thus deep networks are more optimisable compared to shallower ones. Interestingly, minima that attain a low value of the loss function (more negative $\mathcal{E}/p$) are further apart, yet increasing network depth brings even those minima closer together in weight space. 

\emph{Width of minima}: Having shown that there are less minima in deep networks and the minima are closer together, we turn to examine how the width of minima varies with the value of loss function that it attains. To measure basin volume at minima $\mathbf{W}_q$, we consider the entropy $S(\mathbf{W}_q)= -\log \det(\operatorname{Hess}(\mathcal{L}(\mathbf{W}_q)))$, with $\operatorname{Hess}$ being the Hessian matrix \cite{das2016energy,ballard2016energy,zhang2018energy}. Within the harmonic approximation, larger entropy corresponds to larger basin volume. Intuitively, if wider minima are also deeper, then the function is easy to optimise, whereas functions with deep and narrow minima are difficult to optimise. 

The expected entropy of the Hessian of the minima of loss function that takes value $\Lambda \mathcal E$ satisfies asymptotically 
\begin{equation}
\begin{split}
&\mathbb{E}\left( S(\operatorname{Hess}\mathcal L) \vert \Lambda \mathcal E \right)\simeq \\
& -(\Lambda-1)\log\left(p \right) + \tfrac{\Lambda-1}{2}  \log\left(\tfrac{\Lambda}{2(\Lambda-1)H(H-1)}\right) \\
&- \tfrac{\Lambda-1}{\pi} \int_{-\sqrt{2}}^{\sqrt{2}} \log\left\lvert \sigma\sqrt{\tfrac{ \Lambda}{\Lambda-1}} \tfrac{\mathcal E}{p}-t \right\rvert \sqrt{2-t^2} \ dt.
\label{eq:expentrop}
\end{split}
\end{equation}

\emph{Proof}: We start by studying a small energy interval $E=(\mathcal E-\varepsilon,\mathcal E+\varepsilon)$ around some energy $\mathcal E$  where we assume that the auxiliary interval $G=\sigma \sqrt{\tfrac{\Lambda}{\Lambda-1}} E/p$ is contained in $ (-\infty,-\sqrt{2}]$, as minima of the loss function and the spin glass Hamiltonian are known to appear at low energies for large values of $\Lambda$ \cite{auffinger2013random}. 

Let $M_{\mathcal{H}_{\Lambda}}(\Lambda E/p)$ be the event that the Hamiltonian possesses a minimum at some energy in the interval $\Lambda E/p.$ We are interested in finding the expected entropy at those points. We first rewrite this conditional expectation in terms of an auxiliary random variable $X=\tfrac{\sigma\mathcal H_{\Lambda}}{\sqrt{\Lambda(\Lambda-1)}} $ and a GOE matrix $M^{\Lambda-1}$ of size $\Lambda-1$ using the tower property and the probability distribution of the spin glass Hessian \cite[Lemma $1.1$]{auffinger2013complexity}
\begin{equation}
\begin{split}
\label{eq:red1}
&\mathbb{E}\left( S(\operatorname{Hess}\mathcal H_{\Lambda}) \vert M_{\mathcal{H}_{\Lambda}}(\Lambda E/p) \right) \\
&=\frac{\mathbb{E}\left(\mathbb{E}\left( S(\operatorname{Hess}\mathcal H_{\Lambda})1_{M_{\mathcal{H}_{\Lambda}}(\Lambda E/p)} \vert \left\{\mathcal H_{\Lambda} \right\} \right)\right)}{ \mathbb{E}\left( \mathbb{P}\left(M^{\Lambda-1} \ge X, X \in G \vert \left\{ X \right\} \right) \right)}.
\end{split}
\end{equation}
We now consider the asymptotic behaviour of the numerator and denominator separately for large $\Lambda$. The distribution of the Hessian of $\mathcal H_{\Lambda}$\cite[Lemma $1.1$]{auffinger2013random} allows us to express the numerator in terms of an auxiliary function $f_{\beta}(t)=\sqrt{\tfrac{\Lambda-1}{2\pi \sigma^2}} \int_{G} e^{-\frac{\mathcal E^2(\Lambda-1)}{2\sigma^2}} \log \left\lvert t-x \right\rvert \ dx$. Using the Wigner semicircle law, 
\begin{equation} 
\begin{split}
\label{eq: intexp}
&\mathbb{E}\left(\mathbb{E}\left( S(\operatorname{Hess}\mathcal H_{\Lambda})1_{M_{\mathcal{H}_{\Lambda}}(\Lambda E/p)} \vert \left\{\mathcal H_{\Lambda} \right\} \right)\right) \\
&\simeq - \tfrac{\Lambda-1}{\pi}  \int_{-\sqrt 2}^{\sqrt 2} f_{-\sqrt{2}}(t) \sqrt{2-t^2}  \ dt \\
&+\tfrac{\Lambda-1}{2}\log\left(\tfrac{\Lambda}{ 2(\Lambda-1)H(H-1)} \right)\mathbb{P}\left(  M_{\mathcal{H}_{\Lambda}}(\Lambda E/p)\right).
\end{split}
\end{equation}
For the denominator in \eqref{eq:red1}, we use the probability distribution of $X$ and that the lowest eigenvalue of the random matrix $M^{\Lambda-1}$ concentrates at the lower end $-\sqrt{2}$ of the semicircle distribution for $\Lambda$ large \cite[Theorem 1]{LR10}. Hence, it follows that $ \mathbb{E}\left( \mathbb{P}\left(M^{\Lambda-1} \ge X, X \in G \vert \left\{X\right\}\right) \right) =  \sqrt{\tfrac{\Lambda-1}{2\pi \sigma^2}} \int_{G}  e^{-\frac{t^2(\Lambda-1)}{2\sigma^2}}  \ dt.$
Having obtained asymptotic expressions for both the numerator and denominator in \eqref{eq:red1}, we take the limit $\varepsilon \downarrow 0$ such that the energy interval $E$ shrinks down to a single energy value $\mathcal E$ such that \eqref{eq:expentrop} follows immediately. 

\begin{figure}
  \centering
   \includegraphics[width=7cm]{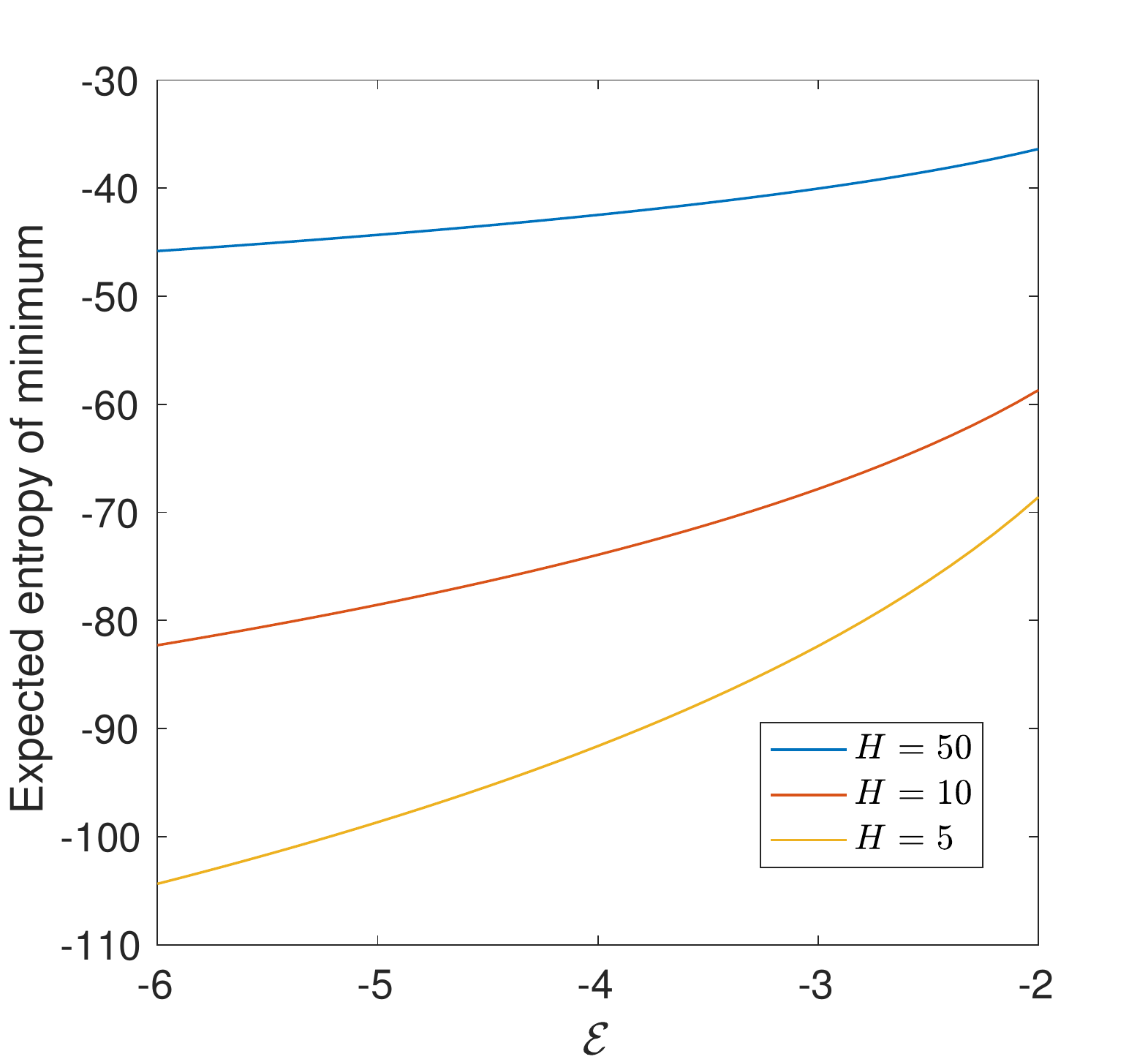} 
   \caption{Energy-entropy competition is eased by increasing network depth. The expected entropy at minima of the loss function as a function of minima depth for $N_e=400$ network parameters and $p=0.8$. }
\label{fig:Netarch2}
\end{figure}

Figure \ref{fig:Netarch2} shows that the lower in loss function that the minima attains, the narrower it is, thus there is an ``energy-entropy'' competition. The existence of energy-entropy competition is non-trivial and unlike many atomic cluster systems analysed in the literature \cite{doye1998thermodynamics,doye2005characterizing,massen2007power}, where the lower minima have larger basins of attraction. However, this competition is smoothened as the number of layer increases. For very deep networks, minima that attains a very low value of loss function has almost the same width as minima that attain a high value of loss function. As such, there is less risk of minimisation algorithms getting trapped in wide but very suboptimal local minima.

\begin{figure}
\includegraphics[scale=0.25]{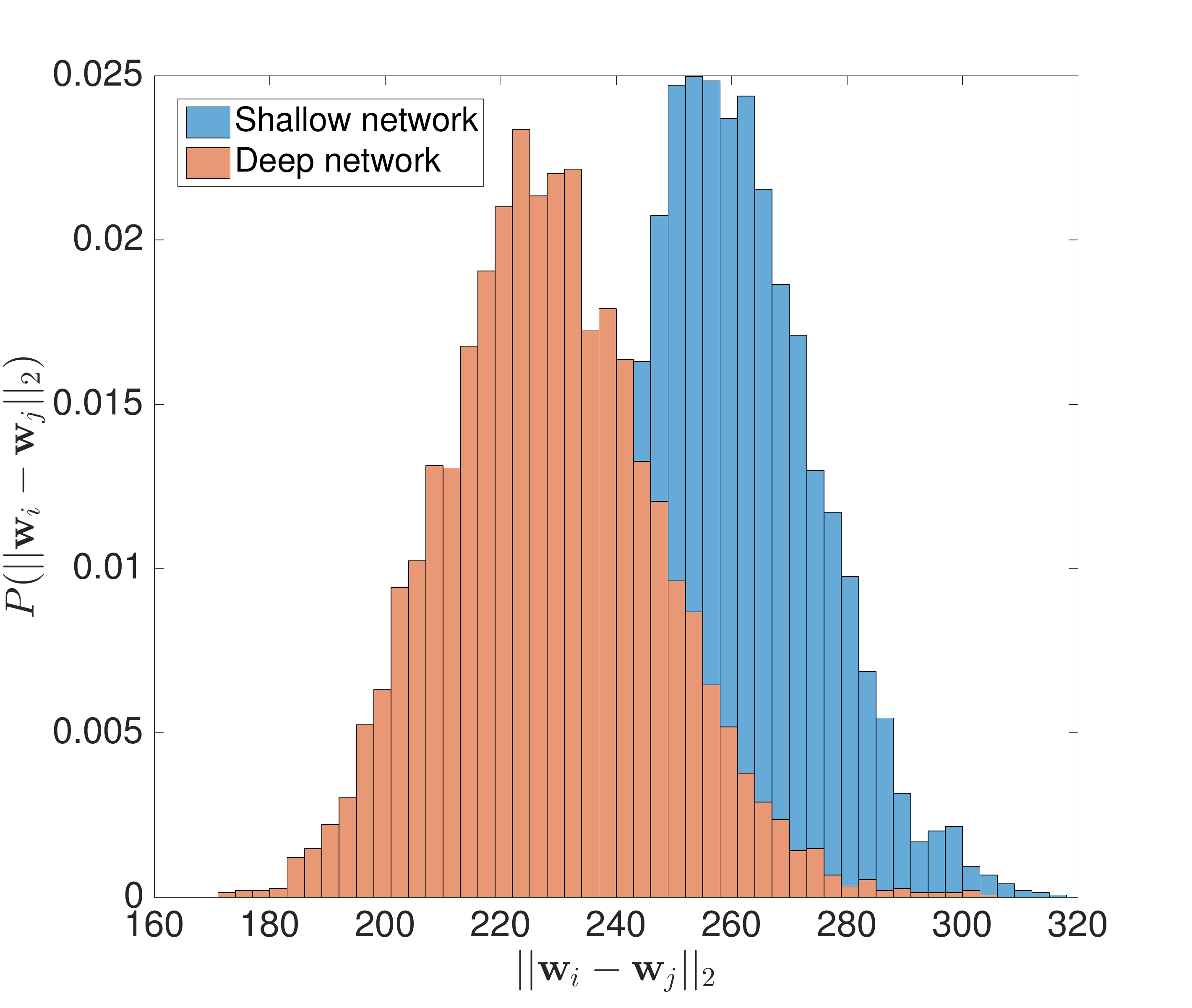} 
\includegraphics[scale=0.27]{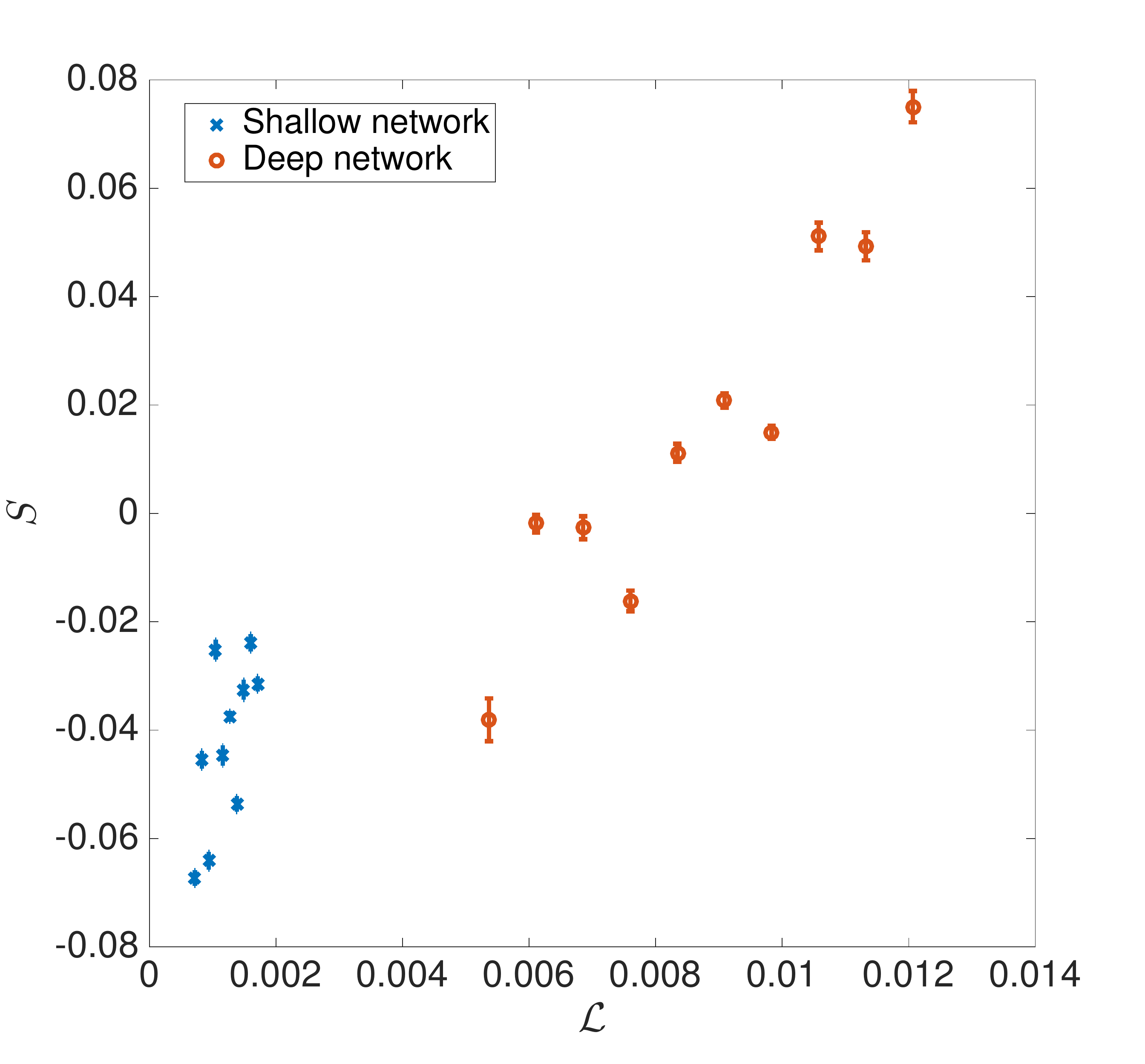} 
 \caption{Numerical experiments agree qualitatively with the analytical predictions. Top: The histogram of distances between minima. Minima in deeper networks are closer together. Bottom: The loss function at minima plotted against the expected entropy of minima. Lower minima are narrower but this energy-entropy tradeoff is less severe for deep networks. The figures are plotted for the ``Boston Housing'' dataset, c.f. \cite{hernandez2015probabilistic, gal2016dropout}. To compute the expected value of entropy, we discretise the distribution of values that that loss function takes into 10 bins.}
\label{exp}
\end{figure}

To verify our analytical results, we consider a classical set of 10 benchmark datasets \cite{hernandez2015probabilistic, gal2016dropout}. Figure \ref{exp} shows the results for one dataset (results for the remaining datasets, shown in the SM, agree with the theory) -- the distance between minima decreases as a function of depth, as shown by the shift in the distribution of pairwise distance between minima, and the tradeoff between minima depth and width is eased. Enumerating the number of critical points is numerically challenging and has only been done for particle systems with relative small number of particles \cite{martiniani2016turning,martiniani2016structural}, thus this is outside the scope of the present study. In the numerical experiments, the input size is 10, the shallow network comprises 2 hidden layers and 22 nodes each and the deep network comprises 6 hidden layers with 22 nodes each, such that the total number of parameters is 726. Further details are discussed in the SM.   


In summary, we derived a series of analytical results showing that deep networks are more optimisable then shallow networks because there are less critical points, the minima are more clustered, and the energy-entropy tradeoff is eased. We verified our analytical results via a set of numerical experiments on classical benchmark datasets in machine learning. Our work sheds light on why deep learning empirically works from the perspective of optimisation, as well as suggests new design principles. For example, the most optimisable machine learning architecture is one where lower minima are also wider, and we speculate that analogies between loss function and energy landscape of atomic systems \cite{doye1998thermodynamics,doye2005characterizing,massen2007power} holds the key to engineering such architectures.

\begin{acknowledgments}
This work was supported by the EPSRC grant EP/L016516/1 for the University of Cambridge CDT, the CCA (S.B.). AAL acknowledges support from the Winton Programme for the Physics of Sustainability.
\end{acknowledgments}

\bibliography{references} 
\end{document}